%% file: main.tex
\documentclass[sigconf,nonacm]{acmart}
\usepackage{tikz}
\usepackage{amsmath}
\usepackage{makecell} 
\usepackage{multirow}
\usepackage{booktabs}
\usepackage{xspace}
\usepackage{enumitem}
\usepackage[normalem]{ulem}

\usepackage{tabularx}

\usepackage{url}
\newcommand\longurl[1]{\mathchardef\UrlBreakPenalty=100
\mathchardef\UrlBigBreakPenalty=100\url{#1}}

\usepackage{tcolorbox}  

\renewcommand{\paragraph}[1]{\vspace{5pt}\noindent\emph{#1.}}

\usepackage{listings}
\usepackage[inkscapelatex=false]{svg} 
\acmConference[The Web Conference '26]{}{April 13-17, 2026}{Dubai, UAE}
\acmISBN{978-1-4503-XXXX-X/2018/06}

\newcommand{\twolinecellcenter}[2][c]{%
  \begin{tabular}[#1]{@{}c@{}}#2\end{tabular}}

\begin{document}

\title{Unpacking .zip: A First Look at Domain and File Name Confusion}


\author{Predrag Despotovic}
\email{despotop@oregonstate.edu}
\affiliation{%
  \institution{Oregon State University}
  \city{Corvallis}
  \state{Oregon}
  \country{USA}
}

\author{Pranab Mishra}
\email{mishrap@oregonstate.edu}
\affiliation{%
  \institution{Oregon State University}
  \city{Corvallis}
  \state{Oregon}
  \country{USA}
}

\author{Kevin Rossel}
\email{rosselk@oregonstate.edu}
\affiliation{%
  \institution{Oregon State University}
  \city{Corvallis}
  \state{Oregon}
  \country{USA}
}

\author{Athanasios Avgetidis}
\email{avgetidis@gatech.edu}
\affiliation{%
  \institution{Georgia Institute of Technology}
  \city{Atlanta}
  \state{Georgia}
  \country{USA}
}

\author{Zane Ma}
\email{zane.ma@oregonstate.edu}
\affiliation{%
  \institution{Oregon State University}
  \city{Corvallis}
  \state{Oregon}
  \country{USA}
}

\renewcommand{\shortauthors}{Despotovic et al.}


\input{sections/00-abstract}

\maketitle

\input{sections/01-introduction}

\input{sections/02-background}

\input{sections/04-filesquatting_tlds}
\input{sections/05-client-testing}
\input{sections/06-honeypots}
\input{sections/07-discussion}

%



\bibliographystyle{plain}
\bibliography{ref}

\appendix

\input{sections/09-appendix}

\end{document}

%% file: sections/00-abstract.tex
\begin{abstract}
The namespace for filenames and DNS names has overlapped since the introduction of DNS in 1985: \texttt{.com} was the original binary format used for DOS and CP/M systems. Recently the introduction of gTLDs such as \texttt{.zip} and \texttt{.mov}, coupled with the growing prevalence of web resources, has ignited new concerns about potential issues related to DNS and filename confusion. 
Thus far, the discourse on DNS/filename confusion has been piecemeal and hypothetical, making it unclear what, if any, security concerns credibly exist.
To address this gap, we provide the first enumeration of how DNS/filename confusion can be abused. 
We then perform the first empirical case studies of DNS/filename confusion in the wild, which highlights suspected confusion across a wide range of software.
Finally, based on our preliminary findings, we provide suggestions and guidance for future research on this topic.

\end{abstract}

%% file: sections/01-introduction.tex
\section{Introduction}

In 2023, Google Registry announced eight new DNS top-level-domains (TLDs) available for public registration, including \texttt{.zip}. Due to the fact that these new TLDs are also common filename extensions, the announcement provoked a flurry of commentary on potential security concerns and non-concerns related to collisions in the namespace for filenames and DNS names~\cite{new-tlds-not-bad-actually,kaspersky-beware-zip-mov,talos-info-leakage,phishing-attacks-using-zip,trend-micro-concerns,wired-new-tld-risks}. When a user or software encounters the string \texttt{document[.]zip}\footnote{The ``[.]'' notation is used to prevent PDF-reader auto-linking to potentially harmful domains.}, this could refer to either a compressed archive file stored locally on the computer, or a website accessible at \url{http://document[.]zip}. This ambiguity creates a namespace collision where the same string can have two entirely different meanings depending on context. While this overlap has existed since 1985 when \texttt{.com} served as both a DOS executable extension and a top-level domain, the 2023 introduction of gTLDs including \texttt{.zip} and \texttt{.mov} by Google Registry has amplified these concerns. 

Unfortunately, the analysis of potential namespace confusion concerns has been piecemeal and theoretical to date and has not thoroughly answered questions such as: ``What harms are possible if the file \texttt{file[.]zip} is mistakenly interpreted as a DNS name, or vice versa? How realistic and frequent are these issues?'' In this paper, we first expand on limited prior work by enumerating all possible attack vectors that arise from name confusion, including novel scenarios not previously considered. Our framework reveals that harms primarily arise when filenames are mistakenly interpreted as DNS names, while the inverse scenario has limited security impacts. 

To go beyond theoretical security concerns, we perform the first empirical study of DNS and filename namespace confusion. We begin by identifying potential software that is likely to encounter ambiguous file and domain names, which consists largely of messaging software that may automatically hyperlink text that appears to be a domain name. After testing different filename and domain name scenarios, we find evidence of client confusion when handling ambiguous domains/filenames. In particular, we identified and ethically reported vulnerable implementations that parse an ambiguous name differently when auto-linking and generating link previews, creating opportunites for user deception. 

We complement our client testing with a honeypot experiment to gain a server-side perspective of domain and filename confusion. We collected seven months of network traffic across six \texttt{.zip} domains, which were chosen based on the popularity of effective top level domains (eTLDs) and filenames, as well as NXDOMAIN errors from passive DNS telemetry. We are able to attribute thousands of requests to previews/clicks from fingerprinted client software, validating our initial findings, but a majority of honeypot traffic remains unattributed, indicating that file and domain name confusion is widespread and further research is needed. 

Overall, this paper introduces the full scope of DNS/filename collision concerns and presents the first empirical evidence that such abuses are possible in practice. We hope that this work informs future implementations of link parsing logic and prevents the unsafe auto-linking behavior that we observed in popular messaging platforms.

%% file: sections/02-background.tex
\section{Background}

\begin{table*}[h]
    \centering
    \begin{tabularx}{\linewidth}{cccX}
    \toprule
    \textbf{\twolinecellcenter{Intended\\Resource}} & \textbf{\twolinecellcenter{Actual\\Resource}} & \textbf{Example} & 
    \textbf{Threats} \\
    \midrule
    Local file & DNS name & \texttt{file[.]zip} & Info.\ leakage, drive-by download, browser exploit, social engineering.\\
    Remote file & DNS name & \texttt{box[.]com@file[.]zip} & Info.\ leakage, malicious download, browser exploit, social engineering. \\
    DNS name & Local file & \texttt{command[.]com}~\cite{wiki:command-com} & Limited, unless malicious/exploitable file exist locally. \\
    DNS name & Remote file & -- & N/A. Intended DNS name cannot contain misleading remote file URL due to limited DNS namespace.\\
    \bottomrule
    \end{tabularx}
    \caption{Threats posed by file and DNS name confusion.}
    \label{tab:threats}
\end{table*}

This work examines two independent namespaces: DNS names and filenames. DNS names are hierarchical. At the top of the hierarchy (ignoring the root zone which has no name) is the top-level domain (TLD), operated by a DNS registry. Below it is the second-level domain (2LD), such as \texttt{foo} in \texttt{foo.com}, preceded by zero or more subdomains. Some TLD registries reserve second-level domains, so the effective TLD (eTLD) is defined as the lowest non-registrable label (e.g., \texttt{co.uk}), and the effective 2LD (e2LD) is the first registrable label beneath it (e.g., \texttt{foo.co.uk}). The DNS namespace therefore consists of all eTLDs, arbitrary e2LDs, and subdomains, with labels restricted to alphanumeric characters and hyphens, a maximum of 63~bytes per label, and 255~bytes for the full domain name~\cite{dns-concepts-rfc,dns-spec-rfc}. The namespace continues to expand as ICANN delegates new TLDs to registries~\cite{icann-gtld-application}.

In most popular file systems, filenames have minimal technical restriction, since they can contain nearly all characters besides a handful of system-specific special characters (e.g., \texttt{:} and \texttt{/} in Mac’s HFS+) up to a maximum filename length [19]. However, in practice, many files contain a ``.'' character followed by a file extension as the filename suffix, which allows users and software to predict a file's type and handle it appropriately. Since most modern file systems allow all permitted characters for DNS names and have a similar 255 character maximum, the namespace for filenames encompasses all DNS names. Whether a given name should be interpreted as a filename or DNS name depends on context---the string \texttt{report[.]zip} may denote a compressed archive or a domain serving arbitrary content, with no explicit marker distinguishing the two. 

DNS names have intersected with common file extensions since their inception. The first public top-level domains (TLDs) announced in 1984~\cite{first-tlds-rfc} included \texttt{.com}, which was a standard extension for executable files on early operating systems, including CP/M and DOS~\cite{wiki:com-file}. ccTLDs such as Paraguay’s \texttt{.py} and St.\ Helena’s \texttt{.sh}, which overlap with Python and shell script files, respectively, have been in use since the 1990s. Most recently, in 2023\footnote{Google became the registry for these TLDs in 2014, but initially prohibited public registration.} Google made its \texttt{.zip} and \texttt{.mov} domains publicly registrable~\cite{google-new-tlds}, sparking a new wave of concerns around DNS and filename confusion.

\subsection{Threat model}

Our threat model (Table~\ref{tab:threats} enumerates all possible confusions between some intended resource and the actual resource it is confused for, which can be a local file, a remote file (as indicated by URL) or a DNS name.

When the intended resource is a local file that is confused for a DNS name, DNS resolution occurs, often followed by a subsequent HTTP request. Even if the target domain is not registered, DNS registries can observe most mistaken resolutions, taking into account the effects of DNS caching. This means Google can observe mistaken \texttt{.zip} lookups and infer files that are present on a system or mentioned via messaging applications. If the domain is registered, the e2LD owner can observe lookups made to that domain or any of its subdomains, and respond to follow-up HTTP requests with potentially malicious responses such as drive-by downloads or exploits for vulnerable browsers. Social engineering may also be effective, since some email providers (e.g., Microsoft Outlook) have introduced ``filename links'' that obfuscate the URL host and display only a filename that links to a cloud-based document (e.g., a Word document in Sharepoint). As such, users might expect that clicking on \texttt{important[.]zip} will open a web-based file.

A second case of confusing a file for a DNS name arises with remote file URLs. Using proven URL manipulations~\cite{reynolds:20:url-identity-confusion}, adversaries can craft URLs that mislead a user into thinking they are accessing a file from a trusted source. For instance, \texttt{https://box[.]com@file[.]zip} can lead to a user-consented, but mistaken, file download, since all characters before the ``@'' are not considered part of the hostname. In Section~\ref{sec:client-testing}, we find that some messaging software mistakenly amplifies this user confusion by improperly previewing the incorrect \texttt{box.com} domain.

When the intended resource is a DNS name that is confused for a file, the possibilities for exploitation are limited. Due to DNS naming limitations that prevent the use of \texttt{:} and \texttt{/} characters, a DNS name cannot be mistaken for a remote file; the only possibility is being mistaken for a local file. In this case, a local file is opened or run instead of a DNS lookup, which limits attacks to restricted living-off-the-land (LotL) techniques~\cite{lotl-attacks} that rely on malicious or exploitable files (e.g., opening an insecure port or overwhelming system resources) already present on the system.

Our threat model assumes three main types of adversaries: 1) passive observers of DNS data, such as TLD registries, 2) domain registrants who can register \texttt{.zip} domains and serve malicious content, and 3) phishing adversaries who craft messages, posts, or URLs to trick users into clicking strings that will be misinterpreted, or to trigger auto-linking in software where confusion occurs.

\subsection{Confusion scenarios}

\begin{figure}[t]
    \centering
    \includegraphics[width=1\columnwidth]{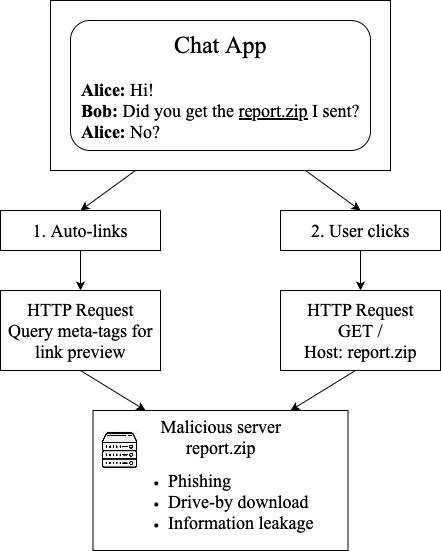}
    \caption{Conceptual threat model of namespace confusion between filenames and domains.}
    \label{fig:dns-http-ratio}
\end{figure}

\paragraph{Auto-linking} A potent scenario for confusion occurs when filenames are automatically hyperlinked by applications such as messaging clients, document editors, or email clients. A user who shares the string \texttt{report[.]zip} may see software convert it into a clickable link pointing to the URL \texttt{http://report[.]zip}. When the recipient clicks this link, the URL is loaded in a web browser that performs a DNS resolution followed by an HTTP request, treating the filename as a domain. If the domain is registered, the registrant can serve arbitrary content, including phishing pages or drive-by downloads. Even when the domain is unregistered, the DNS query itself leaks information about local filenames or message content to any passive DNS listener, e.g., a TLD registry.

\paragraph{Link Preview} Many modern platforms generate link previews or attempt to resolve text to provide better user experiences. In such contexts, if a filename-like string is recognized, such as \texttt{backup[.]zip}, it can trigger automated DNS lookups without any explicit user action. Registries and domain owners observing these requests gain visibility into filenames that may exist on user systems or within private communications. This confusion exposes an information-leak channel, where passive adversaries can gain insight into message content simply because an application attempted to enrich the user interface.

\paragraph{URL Deception} A third class of confusion arises from adversary-crafted URLs that embed filename-like strings in ways that obscure the actual host component. An attacker can exploit parsing rules by constructing a URL such as \texttt{https://box[.]com@file[.]zip}, where the \texttt{box[.]com} part will be ignored and, in reality, \texttt{file[.]zip} will be resolved. In practice, the browser interprets the portion after the ``@'' symbol as the authoritative domain, resulting in navigation to \texttt{file.zip}. The attacker who controls this domain can then deliver malicious payloads. Unlike auto-linking or preview leakage, which rely on inadvertent conversions of filenames, this scenario reflects deliberate adversarial use of namespace overlap to induce user consent to unsafe actions.

%% file: sections/04-filesquatting_tlds.tex
\begin{figure}
    \centering
    \includegraphics[width=\linewidth]{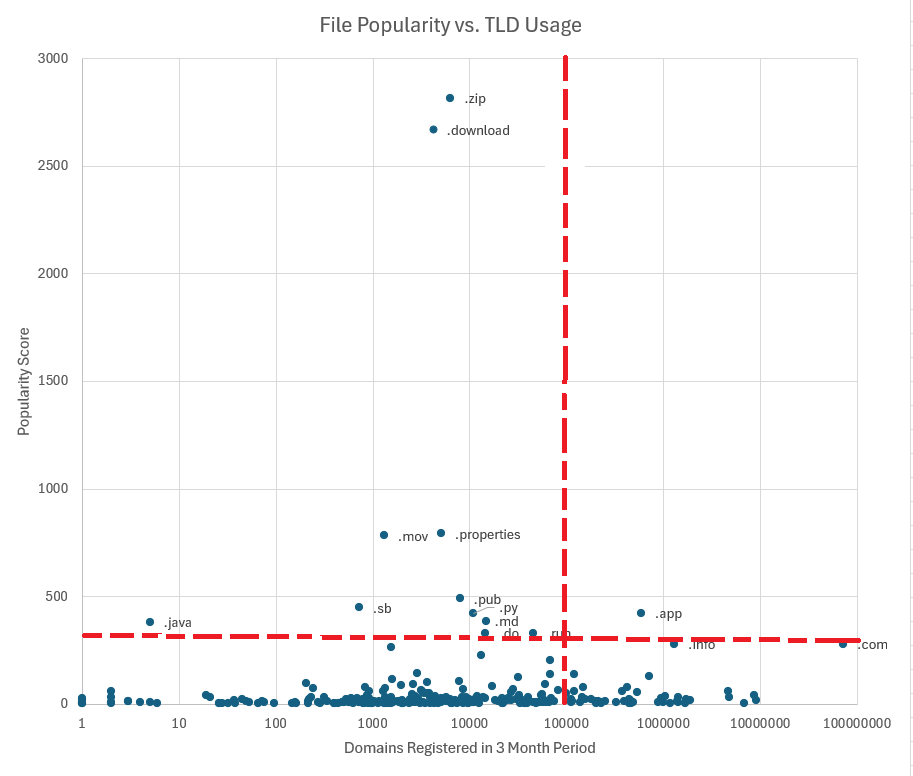}
    \caption{TLDs with likely filename confusion and relatively low DNS usage.}
    \label{fig:candidates}
\end{figure}

\begin{table*}[]
    \centering
    \begin{tabular}{ccc}
    \toprule
    \textbf{Name} & \textbf{\# Lookups} & \textbf{Honeypotted?}\\
    \midrule
      \texttt{tvmt11-flex-ota-2023\_04\_14-10\_19\_39-signed.zip} & 15,923,035 & No \\
       \texttt{gitlab.zip} & 331,810 & Yes \\ 
       \texttt{wallboard-generic-oreo-ota-2022\_02\_17-10\_54\_31-signed.zip} & 59,773 & Yes \\
       \texttt{crowdstrike-hotfix.zip} & 7,709 & Yes \\
       \texttt{downloaddataen.zip} & 3,128 & Yes \\
       \texttt{simcare-v1.1.zip} & 491 & No \\
       \texttt{xray\_ultimate\_1.20.1\_v4.2.5.zip} & 116 & No \\
    \bottomrule
    \end{tabular}
    \caption{Sample of DNS names that resemble filenames and have NXDOMAIN responses.}
    \label{tab:passive-dns}
\end{table*}


\section{Measuring Name Confusion}

Given the range of potential harms and diverse speculation on the impact of DNS and filename confusion, we measured DNS traffic to better quantify real-world instances of name confusion. 

\subsection{Identifying Candidate TLDs}

The first challenge is the lack of ground truth data: every DNS name technically overlaps with a potential filename, so it can be difficult to know which lookups are intentional DNS queries and which are the result of name confusion. To address this issue, we first isolated our measurement to the most likely TLDs to experience detectable filename and DNS name confusion: TLDs that are both commonly used file extensions and relatively unpopular DNS TLDs. 

To approximate file extension popularity, we used \texttt{fileinfo.com}, a file format catalog that accepts user ratings for file extension popularity. While the accuracy of the aggregate user ratings could be manipulated, we are unaware of prior work or a better method for estimating file extension popularity. For TLD popularity, we monitored Certificate Transparency (CT)~\cite{ct-rfc} for a three-month period from September to December 2024 and extracted all DNS names and their eTLDs. CT represents domains that are not just registered but also in use as an HTTPS server. 

Figure~\ref{fig:candidates} shows the popular file extensions (popularity score > 350) that are unpopular DNS TLDs (fewer than 100K domains observed) for identifying candidate TLDs. In total, we found eleven TLDs---\texttt{do, download, java, md, mov, properties, pub, py, run, sb, zip}---that are likely to contain evidence of DNS / filename confusion. Due to resource constraints---some TLDs such as \texttt{.download} charge premium fees for many 2LDs, and country-code TLDs such as \texttt{.py} and \texttt{.sb} restrict registration---we decided to focus on the most popular file extension \texttt{.zip} for this study, which had relatively affordable domain registration fees for our subsequent honeypot analysis.

\subsection{Detecting Potential Name Confusion}

To detect potential name confusion, we looked for the candidate TLDs within a passive DNS dataset containing anonymous, aggregated daily counts of DNS query responses within a large North American ISP from 2024-01 to 2024-10. Rather than look at all queries for candidate TLDs, we instead focus on NXDOMAIN error responses that indicate a DNS lookup for an unregistered DNS name\footnote{In some cases, NXDOMAIN may mean a domain is unavailable for some reason, not necessarily unregistered}. Our insight is that DNS lookups for non-existent \texttt{.zip} (and other) domains can highlight filenames that are mistakenly looked up in DNS, whereas registered DNS names are often legitimate web resources. We manually identified domains that most resembled file names, and a sample of results can be found in Table~\ref{tab:passive-dns}.

\subsection{Honeypot Setup}

To best characterize potential cases of local file and DNS name confusion, we set up a honeypot experiment with the following domains:
\begin{sloppypar}
\begin{itemize}[leftmargin=*]
    \item \textbf{Control domains}: We registered two control domains \texttt{metadata.rodeo} and \texttt{manifest.observer} that have TLDs that do not overlap with any known filename extension. 
    \item \textbf{Simple filenames}: We registered \texttt{metadata.zip} and \texttt{manifest.zip}, which have simple filenames that may overlap with local files. 
    \item \textbf{Popular NXDOMAINs}: We registered \texttt{gitlab.zip}, \texttt{crowdstrike-hotfix.zip}, \texttt{downloaddataen.zip},  and \texttt{wallboard-[...].zip}. 
\end{itemize}
\end{sloppypar}
For all eight domains, the honeypot tracked all DNS traffic to the authoritative nameserver, running \texttt{gdnsd}~\cite{gdnsd}, as well as all traffic to a Caddy~\cite{caddy} webserver. We also collected decrypted HTTP logs for all HTTP/HTTPS traffic sent to the webserver. 

%% file: sections/05-client-testing.tex
\section{Client Testing}
\label{sec:client-testing}

As a starting point for our analysis of file and domain name confusion, 
we tested the behavior of common messaging and email software in order to 
confidently identify known instances of name confusion. 
Based on prior work on auto-linking and link previews~\cite{link-previews}, we selected the top four most popular instant messaging apps (excluding Skype, which is no longer active) and also included Telegram, which is a popular security focused application. Additionally, we tested two popular email clients, Google Gmail in the web browser and Microsoft Outlook. The following versions were tested: WhatsApp iPad version 2.25.24.77, Telegram MacBook version 11.9.272031, Telegram iPad version 12, Telegram iPhone version 12, Snapchat iPad version 13.59.0, Instagram iPad version 398.1.0, Messenger iPad version 525.0.0.24.106, Outlook iPad version 4.2534.1, Gmail in Safari on iPadOS 18.6.2 and macOS Sequoia 15.6.1.

\newcommand{\tikzxmark}{%
\tikz[scale=0.23] {
    \draw[line width=0.7,line cap=round] (0,0) to [bend left=6] (1,1);
    \draw[line width=0.7,line cap=round] (0.2,0.95) to [bend right=3] (0.8,0.05);
}}

\begin{table}
    \centering
    \begin{tabular}{lccccc}
    \toprule
    & \makecell{\textbf{Test \#1} \\ \scriptsize{General AL}} 
    & \makecell{\textbf{Test \#2} \\ \scriptsize{List of TLDs}} 
    & \makecell{\textbf{Test \#3} \\ \scriptsize{Remote URL}} 
    & \makecell{\textbf{Test \#4} \\ \scriptsize{File TLDs}} 
    & \makecell{\textbf{Test \#5} \\ \scriptsize{TLD cutoff}} \\
    
    \midrule
    WhatsApp & \checkmark & \checkmark & \checkmark & \checkmark & \checkmark \\ 
    Telegram & \checkmark & \checkmark & \tikzxmark & \tikzxmark & - \\ 
    Snapchat & \checkmark & \checkmark & \checkmark & \checkmark & \checkmark \\ 
    Instagram & \checkmark & \checkmark & \tikzxmark & \checkmark & \checkmark \\ 
    Messenger & \checkmark & \checkmark & \tikzxmark & \tikzxmark & \checkmark \\
    Gmail & \checkmark & \checkmark & \checkmark & \checkmark & \checkmark \\
    Outlook & \checkmark & \checkmark & \checkmark & \checkmark & \checkmark \\
    
    \bottomrule
    \end{tabular}
    \caption{\textbf{Client testing}---A \checkmark\ indicates \textit{safe} client behavior, while an X denotes \textit{improper or potentially unsafe} behavior, such as linking popular file extensions or incorrectly parsing deceptive remote file URLs.}
    \label{tab:test-results}
\end{table}

\subsection{Test suite}
\label{sec:test-suite}
We conducted five tests, each designed to evaluate how popular messaging and email clients interpret and handle potential confusion. 
For our test suite, we adopt a binary classification of outcomes: safe and potentially unsafe. An outcome is considered safe when the client’s behavior aligns with security-preserving expectations (i.e., non-existent TLDs are not auto-linked, previews are generated only for valid and intended domains, and no unintended protocol interpretations occur). In contrast, we label an outcome as potentially unsafe when the client exhibits behavior that, while not necessarily exploitable in isolation, increases the risk of user confusion, information leakage, or inadvertent navigation. Examples include auto-linking syntactically valid but non-existent TLDs or generating previews for the wrong domain.

\begin{enumerate}[label=\textbf{Test \#\arabic*:}, leftmargin=*, itemindent=34pt, listparindent=0pt, align=left]
  
    \item \textbf{Baseline auto-linking and preview behavior.}
    The purpose of this test is to determine the general auto-linking (AL) and link preview (LP) capabilities of each client by testing different 2LDs with the three most popular gTLDs by Statista~\cite{statista_tlds}: \texttt{.com}, \texttt{.net}, and \texttt{.org}. We also included \texttt{.gov} and \texttt{.edu} as well known high-trust institutional TLDs.

    \item \textbf{Detection of predefined TLD lists.}
    The goal of this test is to identify whether a client references a list of known TLDs, or if they use a simpler pattern-matching approach. We perform this test with three non-existent TLDs (\texttt{.lx}, \texttt{.lss}, and \texttt{.rfl}), to see if they will be autolinked.

    \item \textbf{Deceptive URL with and w/o protocol prefixes.}
    This test identifies how the client handles remote URLs, with and without the \texttt{http(s)://} prefix. We used \texttt{box[.]com@file[.]zip} as a test domain. Since everything before the @ sign is not considered a part of the hostname, we consider auto-linking and previewing \texttt{file[.]zip}---rather than \texttt{box[.]com}---as safe behavior.

    \item \textbf{File extension TLDs.}
    This test determines how the most popular file extensions, which are also TLDs, are interpreted by the client. Our interest lies mostly in \texttt{.zip} domains, but we also tested whether software works off a blocklist for file-extension TLDs. For this purpose, we used \texttt{.zip}, \texttt{.download}, \texttt{.mov}, \texttt{.properties}, \texttt{.app}, \texttt{.info}, \texttt{.pub}, \texttt{.py}, and \texttt{.md}. We used \texttt{something[.]zip} (registered, but no A record), \texttt{somethinggg[.]zip} (not registered), and \texttt{document[.]zip} (registered, A record exists) for testing purposes. By safe behavior, we consider no auto-linking or previewing as safe.

    \item \textbf{Temporal cutoff in TLD lists.}
    This test was only performed if the client did not auto-link \texttt{.zip} strings. One alternate explanation that we want to account for is a temporal cutoff on the list of known TLDs, rather than a file extension blocklist. For this test we used the following TLDs to identify whether \texttt{.zip} is ignored simply because it is a relatively new TLD, as well as try to spot any other patterns (registration dates in parenthesis): \texttt{.pizza} (Aug 2014), \texttt{.bar} (Feb 2014), \texttt{.ceo} (Dec 2013), \texttt{.xyz} (Feb 2014), \texttt{.dog} (Apr 2015), .bom (May 2015), \texttt{.biz} (Jun 2001) and \texttt{.py} (Sep 1991). 
\end{enumerate}

As part of our client testing, we also sent traffic to our honeypot domains, which allowed us to collect ground-truth traffic and extract HTTP fingerprints for each client and each operation: auto-linking and link preview. In combination with aggressive scanner filtering, we can utilize these fingerprints to help identify specific sources of traffic to our honeypots. We were able to fingerprint multiple clients for both operations, as well as differentiate fingerprints across platforms in some cases (web, Android, iOS). Details about the fingerprints can be found in Appendix~\ref{appendix:fingerprints}. 

\subsection{Test Results}

The first two tests were used to determine the baseline behavior of each client. All of the tested clients passed these tests, indicating that they all implement auto-linking and check valid TLDs, as none of them auto-linked non-existent domains. 

When running Test \#3 without the \texttt{http(s)://} prefix, WhatsApp, Snapchat, Messenger, and Telegram (except on macOS) treated the test string \texttt{box[.]com@file[.]zip} as an email address and auto-linked it to the default email client. Outlook exhibited similar behavior, except that it auto-linked the string to its own compose email feature. Gmail did not auto-link the string, while Telegram on macOS exhibits undetermined behavior by changing focus to a non-default browser. Two applications exhibited vulnerabilities. Specifically, for the input string \texttt{https://box[.]com@file[.]zip}, both Messenger and Instagram correctly linked to \texttt{file[.]zip} but incorrectly displayed a preview for \texttt{box[.]com}. This vulnerability confuses users and facilitates social engineering~\cite{link-previews}. A user may be led to believe that clicking the link will lead to \texttt{box[.]com}, when it actually leads to \texttt{file[.]zip}. This behavior is not limited to \texttt{.zip} domains, but applies to any TLD. The observed behavior suggests that Messenger and Instagram perform auto-linking at the client level, while link previews are generated on the server side using a different URL parser.
We responsibly reported these issues to the Meta Bug Bounty program. Meta has since replied acknowledging the issue, and have deployed a fix.

Test \#4 further indicated distinct server-side interpretation of domain names for Messenger: no domains overlapping with file extensions were auto-linked or previewed, except those using the \texttt{.py} and \texttt{.md} TLDs. These two TLDs were not auto linked, but a link preview was still generated. This behavior also represents a social engineering risk, as the previewed strings could mislead users into believing a file was attached---an example of namespace confusion. Telegram was another client that exhibited a vulnerability, auto-linking every domain name and previewing only domains with the \texttt{.info} TLD. All other clients auto-linked and/or previewed only \texttt{.info} domains, while Gmail also auto-linked \texttt{.app} domains. One important conclusion from Test \#4 is that clients do not appear to have a temporal cutoff of TLDs (since .py predates .info). Instead, they seem to use custom logic to determine which TLDs will be auto-linked and previewed, and which will not. This implies that developers of these applications actively attempted to prevent namespace confusion between TLDs and filenames. However, as we see, this logic is inconsistent across application developers and sometimes vulnerable to abuse.

To verify the assumption that clients do not have a temporal cutoff for auto-linking TLDs (e.g., relying on an outdated library), we tested different TLDs that do not overlap with file extension names. For Test \#5 for all clients, we found that the temporal component is not crucial, e.g. \texttt{.py} isn't auto-linked by any client, but is older than most of the domains we tested.

\begin{tcolorbox}[colback=blue!5!white, colframe=violet!80!black, title=Takeaways] Across a range of popular messaging and email software, we observed widespread auto-linking and link preview generation. While there is some avoidance of common filename extensions when auto-linking, this custom logic differs across vendors, creating opportunities for abuse. Additionally, we observed differences in name parsing for auto-linking and link preview in Messenger and Instagram, presenting an immediate social engineering risk that was responsibly disclosed, acknowledged, and fixed.
\end{tcolorbox}

%% file: sections/06-honeypots.tex
\section{Honeypot Case Studies}

To gain a broader understanding of the software that confuses file names and domain names, 
we ran DNS and HTTP(S) honeypots for several \texttt{.zip} domains. Ideally, we are 
interested in the network traffic that originates from a file/DNS name confusion, but isolating 
this traffic from the profusion of automated bot and scanning traffic is very challenging. 

\subsection{Filtering}

\begin{figure}[t]
    \centering
    \includegraphics[width=\columnwidth]{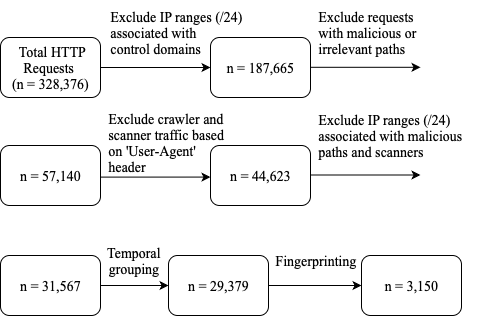}
    \caption{Filtering pipeline for HTTP request dataset. Each stage represents a filtering step applied to remove irrelevant or automated traffic, showing the remaining number of requests ($n$) after each step.}
    \label{fig:filtering-pipeline}
\end{figure}

Over the course of seven months from 2025-02 to 2025-09, we identified 328,376 requests for our six honeypot domains. The two control domains, \texttt{manifest.observer} and \texttt{metadata.rodeo}, had an additional 67,737 requests. 

For the first step of filtering, we leveraged requests to control domains to remove web scanners. Since the control domains are new registrations without file extension TLDs, we assume that almost all of the traffic to the control domains is from web crawlers and scanners. Rather than just filtering out exact-match IP addresses, we took a slightly more aggressive approach and aggregated the scanner IPs to their corresponding /24 subnet representation. We were able to collect 4506 unique /24 CIDR blocks for the control domains, and by filtering out these sub-networks, we were able to remove 140,711 requests, leaving us with 187,665 requests for our honeypot domains.

The manual inspection of the remaining requests showed that a majority of the HTTP requests were directed to paths such as  \texttt{.git/credentials}, \texttt{/config/aws}, \texttt{/credentials.json}, etc., which indicate security scanners or bots trying to exploit possible security vulnerabilities. While profiling the auto-linking and link preview behavior of popular messaging applications, we had noted that such requests were always directed to the root path (\texttt{/}), favicon path (\texttt{/favicon.ico}), and, in a few applications, additional assets such as '\texttt{/apple-touch-icon.png}.' 
Based on this evidence, we removed all requests that queried any path other than these allow-listed paths and were able to filter out an addition 70\% of requests from the previous step, resulting in a remaining set of 57,140 requests.

Our next filter was a set of manually identified HTTP User-Agent headers that we manually identified as bot traffic, including 'GPTBot' 'GoogleBot', 'MJ12bot', 'leakix', etc. We also filtered out requests from programmatic and automated tools like 'curl', 'python-requests', 'Go-http-client/1.1', etc., which are not relevant for our confusion scenarios. In total we collected 21 User-Agent identifiers to exclude from our remaining dataset, removing 12,517 further requests.

Even after removing requests with irrelevant paths and scanning or bot-related User Agents, we observed that the IP addresses responsible for such requests also generated traffic to the root and favicon path. We assume that requests from such IP addresses are also a part of scanning traffic that can be removed. To achieve this, we defined a set of 10 identifiers like '.env', '.git', 'config', '.aws', etc. to group potentially malicious paths and collected the IP addresses that accessed them. In addition, we gathered IP addresses associated with User-Agent containing 21 identifiers identified in the previous step. These IP addresses were aggregated to 3,566 unique /24 subnet blocks. Excluding all requests associated with these IP blocks yielded 31,567 requests.

Our final data cleaning step was a coarse-grained rate-limit filter that removed all IPs making more than 4 requests in a 1 second window. This filter yielded a final dataset of 29,379 HTTP requests to be analyzed. 


\subsection{Fingerprint Results}

\begin{table}
    \centering
    \small
    \begin{tabular}{l l c | l l c}
    \toprule
    \textbf{Client} & \textbf{Action} & \textbf{Count} & \textbf{Client} & \textbf{Action} & \textbf{Count} \\
    \midrule
    \multirow{2}{*}{WhatsApp} & Preview & 11 &     \multirow{2}{*}{Messenger} & Preview & 373 \\
                               & Click   & 0 &                                & Click & 71 \\
    \midrule
    \multirow{2}{*}{Snapchat}  & Preview & 10 &     \multirow{2}{*}{Outlook}   & Preview & N/A \\
                               & Click & 0 &                                & Click & 0 \\
    \midrule
    \multirow{2}{*}{Telegram}  & Preview & 16 &     \multirow{2}{*}{Gmail}     & Preview & N/A \\
                               & Click   & 2,391 &                                & Click & 0 \\
    \midrule
    \multirow{2}{*}{Instagram} & Preview & 244 &     \multirow{2}{*}{\textit{Unknown}}                    & \multirow{2}{*}{\textit{Unknown}} & \multirow{2}{*}{26,229} \\
                               & Click   & 34 & \\
    \bottomrule
    \end{tabular}
    \caption{Detected activity of fingerprinted software}
    \label{tab:fingerprint}
\end{table}

By applying our HTTP fingerprints developed in Section~\ref{sec:test-suite} to the filtered HTTP log dataset, we were able to identify the source for thousands of user clicks and link previews (Table ~\ref{tab:fingerprint}). Clients exhibited distinct behaviors in how they access the honeypot domains. For instance, Telegram generated a large number of Click requests compared to Preview, indicating active user interaction, while Messenger produced a relatively high number of both previews and clicks. One possible reason for Telegram having significantly less Previews than Clicks is that Telegram generates a preview only when the string begins with the HTTP(S) scheme (\texttt{http(s)://example[.]com}) or ends with a slash (\texttt{example[.]com/}), and we would next expect file/DNS confusion scenarios to include either a protocol prefix or slash suffix. We discovered an unexpectedly high number of requests from Vietnam generating Messenger previews; however, since there is no evidence that Meta operates servers in Vietnam, this suggests potentially spoofed User-Agent headers. Instagram also issued multiple preview requests but fewer clicks, indicating that most interactions stopped at preview generation and did not progress to link clicking. In contrast, Outlook and Gmail did not generate any HTTP requests to our domains, consistent with their restrictive link previewing behavior.

We additionally examined the temporal patterns, request frequency distributions, and geographic origins of preview and click requests. To illustrate the results, we present the cumulative distribution function (CDF) of requests per IP that provides insight into traffic intensity and regularity; distribution graphs that show which autonomous the requests originated from, helping identify whether traffic patterns align with known client infrastructure and usage patterns or suggest spoofed or automated sources; the geographic distribution charts highlight where requests appear to come from.

\begin{figure}[t]
    \centering
    \includegraphics[width=1\columnwidth]{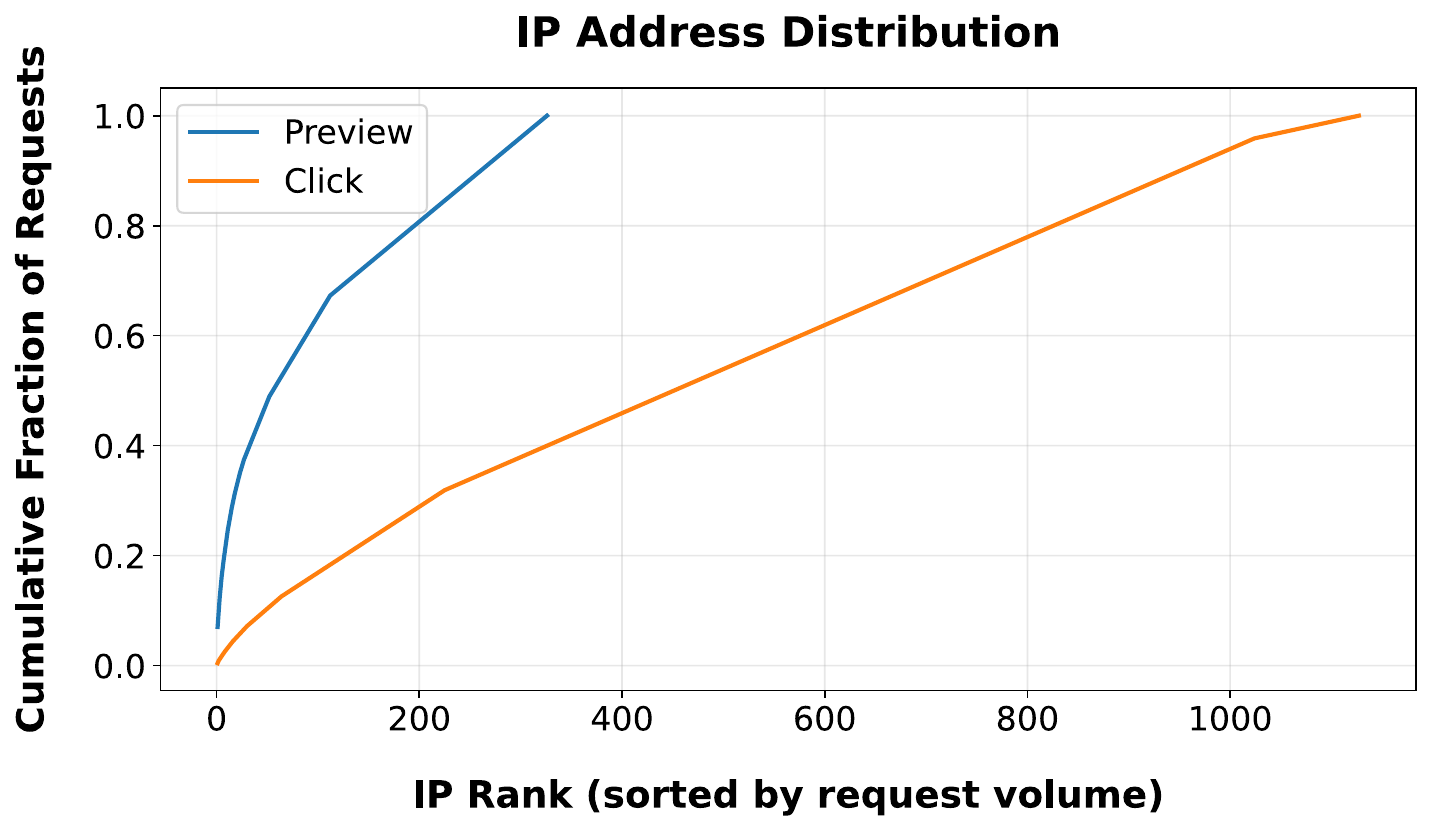}
    \caption{Cumulative distribution function for number of requests per IP address}
    \label{fig:cdf}
\end{figure}

The cumulative distribution function (CDF) shown in Figure~\ref{fig:cdf} of click requests per IP address shows that 70.9\% of IPs made exactly two requests—typically one to the root path and another to \texttt{/favicon.ico}. This pattern aligns with what we observed during manual log inspection: end-users accessing websites through browsers that fetch both the main page content and the favicon for the tab icon. Further inspection revealed that some clients, such as Messenger, occasionally request additional assets such as \texttt{/apple-touch-icon.png}, resulting in more than two requests per IP.
In contrast, most preview requests consist of a single query to the root path, aimed at retrieving metadata tags (e.g., OpenGraph or Twitter Card information) to generate content-rich previews. Some clients also query \texttt{/favicon.ico} during preview generation, while others, such as Messenger, may also request \texttt{/robots.txt}.

\begin{figure*}
    \centering
    \includegraphics[width=2\columnwidth]{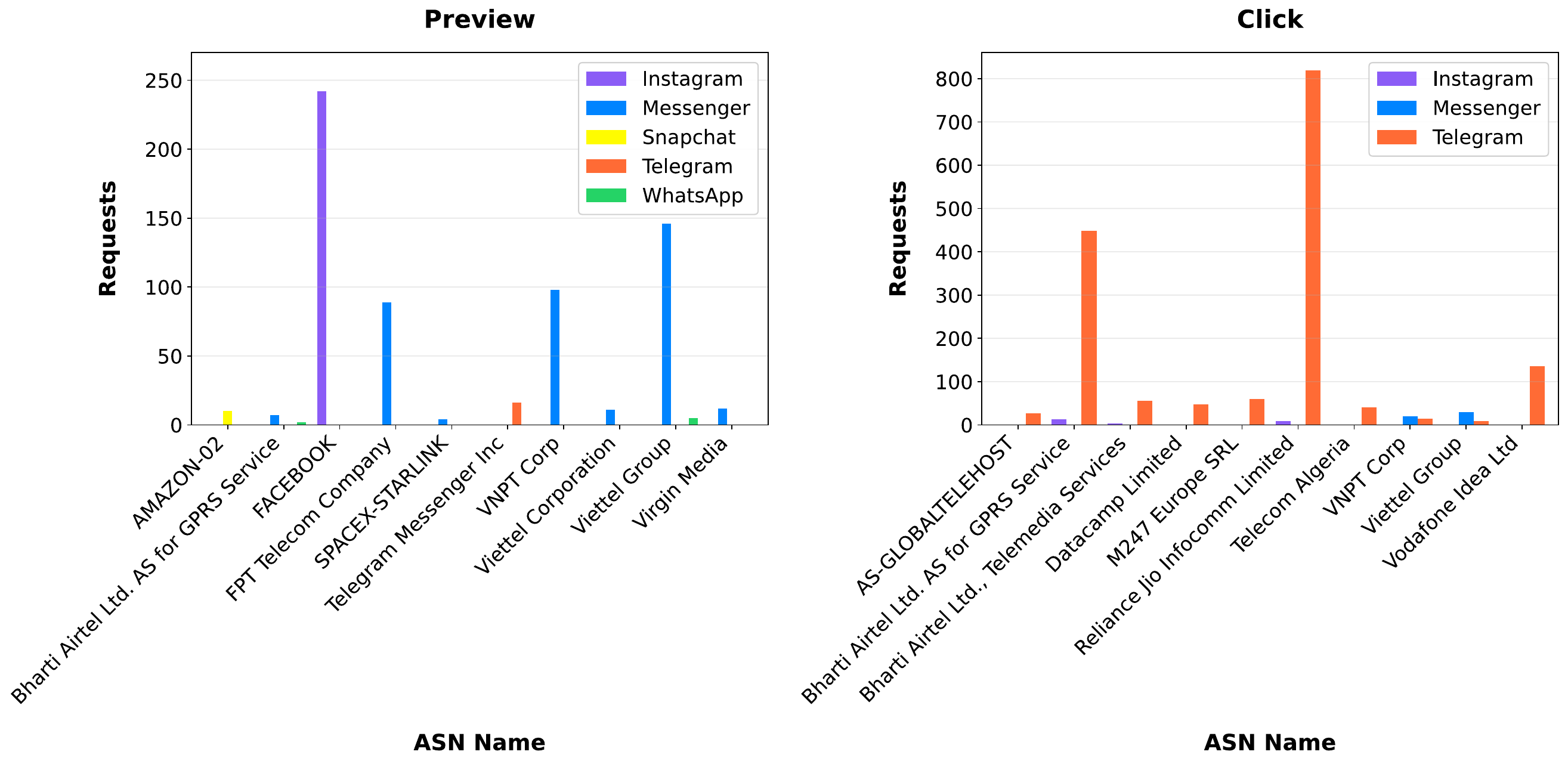}
    \caption{Geographic distribution of preview and click requests}
    \label{fig:asn-distribution}
\end{figure*}

\begin{table}
    \centering
    \small
    \begin{tabular}{l c | l c}
        \toprule
        \textbf{Preview} & & \textbf{Click} & \\
        \toprule
        \textbf{Country} & \textbf{Share} & \textbf{Country} & \textbf{Share} \\
        \midrule
        Vietnam & 53.5\% & India & 65.3\% \\
        United States & 29.4\% & Bangladesh & 4.3\% \\
        United Kingdom & 4.3\% & Vietnam & 3.9\% \\
        Denmark & 3.5\% & United States & 2.5\% \\
        Ireland & 3.4\% & Singapore & 2.4\% \\
        India & 3.2\% & \textit{Others} & \textit{21.6\%} \\
        \textit{Others} & \textit{2.8\%} & & \\
        \bottomrule
    \end{tabular}
    \caption{Geographic distribution of previews and clicks}
    \label{tab:country-distribution}
\end{table}

Table~\ref{tab:country-distribution} and Figure~\ref{fig:asn-distribution} present the network- and country-level origins of preview and click requests, respectively. The ASN distribution shows that preview traffic is dominated by Meta clients (Messenger and Instagram), primarily originating from Facebook’s ASN and several Vietnamese network providers (VNPT Corp, Viettel Group). While preview requests are concentrated within just 20 ASNs, click requests---prevalently from Telegram---span over 200 ASNs across India, Algeria, and the United Kingdom, indicating a far broader network footprint.

The geographic distribution mirrors these patterns. More than half of all preview requests originated from Vietnam, but we found no evidence of Meta infrastructure in that region. This strongly suggests automated activity or spoofed User-Agent headers rather than legitimate client traffic. Click requests, on the other hand, were concentrated in India—consistent with Telegram’s massive user base in the region (over 83.8 million downloads, the highest worldwide), but spread across a wide range of IPs, indicating a diverse set of clients. Together, the ASN and geographic data reinforce the distinction between centralized, likely server-side preview fetching, and more distributed, user-driven client clicks.

\subsection{Other User Agents}

There were a total of 1804 requests with the Referer header set to ``android-app://com.google.android.googlequicksearchbox.'' This header is associated with requests originating from Google's Android application, most likely from the Google search bar widget. Interestingly, all of these requests are for \texttt{metadata.zip} and its root path (\texttt{/}). These requests have occurred daily at irregular intervals each day since August 1, 2025, from geographically distributed IPs and with no common characteristics between requests other than the same value for User-Agent and Referer headers. We assume that these instances represent filename misinterpretation, where a user entered the filename in the search bar widget to search for a file in the file system or where some other application redirected the query to the Google search box of browsers like Chrome.

A total of 594 requests were identified with User-Agent containing ‘MicroMessenger’, which is the built-in browser (WebView) of the popular Chinese social media application WeChat. The requests were distributed across all our honeypot domains and originated from a total of 91 IP addresses with 9 distinct ASNs, all except one belonging to China. However, only five requests were observed for the favicon.ico path (/favicon.ico), while the rest were directed to the root path (/). As there were very few requests for the favicon.ico, there is no strong evidence to indicate link preview generation, in contrast to other messaging platforms that we’ve investigated.


We also found 85 requests from the ``SkypeUriPreview'' User-Agent. These originate from a preview bot that fetches webpage metadata to create a preview of the link when users share URLs in a Skype message. In nearly every case, each IP address generated three simultaneous requests, two directed to the root path (/) and one to the favicon path (/favicon.ico). Interestingly, all of the IP addresses belong to AS8075, which belongs to Microsoft, indicating these requests likely originated from the link preview service of Microsoft, where a filename was mistaken as a valid DNS name.


\begin{tcolorbox}[colback=blue!5!white, colframe=violet!80!black, title=Takeaways]
Definitively identifying DNS and filename confusion is challenging due to high volumes of bot and scanning traffic. However, through rigorous filtering and fingerprinting, we were able to identify organic traffic from link previews and link clicks from messaging apps (e.g., Telegram), social media (e.g., Meta products, Snapchat), and omniboxes (e.g., Android), but most honeypot traffic remains unattributed. 
\end{tcolorbox}

%% file: sections/07-discussion.tex
\section{Discussion}

Our study demonstrates that file and DNS namespace confusion is not a purely theoretical concept, but a measurable, recurring phenomenon that is observable across a range of software. Further study is needed to better estimate the overall prevalence of name confusion across all file and domain names; however, we expect this to be large given that our six honeypot domains found thousands of known lookups while looking at only a tiny silver of the overall overlapping namespace. 


The client test suite revealed that messaging and email platforms exhibit inconsistent approaches to auto-linking and preview generation. Even among reputable mainstream applications, implementations differ in what string is considered a link. Messenger and Instagram, for instance, generated incorrect previews for deceptive URLs such as \texttt{https://box[.]com@file[.]zip}, thereby reinforcing user trust in a misleading and potentially malicious link. Telegram's behavior was more permissive in linking but selective in preview generation, while Gmail and Outlook remained conservative. These discrepancies highlight an absence of shared standards for link parsing and preview generation, allowing subtle namespace ambiguities to manifest as user-facing confusion or inadvertent data exposure. We recommend establishing default-secure standard linking behavior across different apps to reduce user confusion and mistaken queries. 




Finally, as a short-term mitigation strategy, we recommend implementing a universal block-list of TLDs that should not be auto-linked by client software. Given that our analysis revealed confusion requests numbering in the thousands, this approach represents a practical and proportionate safeguard. If subsequent research identifies confusion at a broader scale, a policy-level intervention would be warranted---specifically, a review of ICANN's procedures for approving new TLDs and reassessing existing ones that overlap with common file extensions.

\subsection{Limitations}

This paper only discusses the security concerns of file and DNS name confusion. It does not balance these concerns with the potential utility and benefits of \texttt{.zip} and similar TLDs, which is necessary for a more holistic consideration of the topic. Furthermore, our honeypot deployment covered only six \texttt{.zip} domains, and our results may not generalize to other file extension TLDs such as \texttt{.py} or \texttt{.md} if they have different usage properties. 

Client identification relied on HTTP headers and behavioral fingerprints. Although consistent structural patterns increase confidence, User-Agent spoofing and shared infrastructure could introduce false positives. This is especially relevant for traffic that appears to originate from large providers. Nearly every component of an HTTP request used to identify the source could be manipulated. We believe that our aggressive filtering removes many such cases, but more advanced spoofing can still evade our filters. 


Moreover, while we were able to find patches of likely DNS and filename confusion in the wild, we lacked the data to determine whether such confusion was being actively abused. Future studies with more comprehensive DNS and HTTP data are needed to detect whether and to what degree popular filename extension TLDs are used for abuse based on namespace confusion.



%% file: sections/09-appendix.tex
\newpage
\onecolumn
\section{Software Fingerprints}
\label{appendix:fingerprints}
\begin{table*}[h!]
    \centering
    \small
    \begin{tabularx}{\textwidth}{l l l X}
    \toprule
    \textbf{Application} & \textbf{Platform} & \textbf{Action} & \textbf{Fingerprint Characteristics} \\ 
    \midrule
    
    \multirow{4}{*}{WhatsApp} 
    & \multirow{2}{*}{Android} 
      & Preview & \texttt{User-Agent:} \texttt{WhatsApp/<server-version>} \\
    &  & Click & Not fingerprintable; opens default browser \\
    & \multirow{2}{*}{iPhone} 
      & Preview & \texttt{User-Agent:} \texttt{WhatsApp/<server-version>} \\
    &  & Click & Not fingerprintable; opens default browser \\
    
    \midrule
    \multirow{4}{*}{Telegram}
    & \multirow{2}{*}{Android} 
      & Preview & \texttt{User-Agent:} \texttt{TelegramBot} (like \texttt{TwitterBot)} \\
    &  & Click & \texttt{X-Requested-With} contains “telegram” \\
    & \multirow{2}{*}{iPhone} 
      & Preview & \texttt{User-Agent:} \texttt{TelegramBot} (like \texttt{TwitterBot)} \\
    &  & Click & Not fingerprintable; opens default browser \\
    
    \midrule
    \multirow{4}{*}{Snapchat}
    & \multirow{2}{*}{Android} 
      & Preview & \texttt{User-Agent} contains \texttt{Snap URL Preview Service} \\
    &  & Click & Not fingerprintable; opens default browser \\
    & \multirow{2}{*}{iPhone} 
      & Preview & \texttt{User-Agent} contains \texttt{Snap URL Preview Service} \\
    &  & Click & \texttt{User-Agent} contains \texttt{Snapchat} \\
    
    \midrule
    \multirow{4}{*}{Instagram}
    & \multirow{2}{*}{Android} 
      & Preview & \texttt{User-Agent:} \texttt{facebookexternalhit} \\
    &  & Click & \texttt{User-Agent:} \texttt{Instagram, Android}; \texttt{Referer: https://l[.]instagram[.]com} \\
    & \multirow{2}{*}{iPhone} 
      & Preview & \texttt{User-Agent:} \texttt{facebookexternalhit} \\
    &  & Click & \texttt{User-Agent:} \texttt{Instagram}, device name; \texttt{Referer: https://l[.]instagram[.]com} \\
    
    \midrule
    \multirow{4}{*}{Messenger}
    & \multirow{2}{*}{Android} 
      & Preview & \texttt{User-Agent:} \texttt{WhatsApp/<server-version> <platform>} \\
    &  & Click & \texttt{User-Agent:} Android with Messenger tags; \texttt{Referer:} \texttt{http://m[.]facebook[.]com} \\
    & \multirow{2}{*}{iPhone} 
      & Preview & \texttt{User-Agent:} \texttt{WhatsApp/<server-version> <platform>} \\
    &  & Click & \texttt{User-Agent:} device name with Messenger tags \\
    
    \midrule
    \multirow{4}{*}{Outlook}
    & \multirow{2}{*}{Android} 
      & Preview & No preview \\
    &  & Click & Not fingerprintable; opens default browser \\
    & \multirow{2}{*}{iPhone} 
      & Preview & No preview \\
    &  & Click & Not fingerprintable; opens default browser, preceded by request with no \texttt{User-Agent} (Outlook link safety check) \\
    
    \midrule
    \multirow{4}{*}{Gmail}
    & \multirow{2}{*}{Android} 
      & Preview & No preview \\
    &  & Click & Not tested \\
    & \multirow{2}{*}{iPhone} 
      & Preview & No preview \\
    &  & Click & Not fingerprintable; opens default browser with \texttt{Sec-Fetch-Site: cross-site} \\
    
    \bottomrule
    \end{tabularx}
    \caption{Summary of Request Fingerprints for Messaging Applications -- \textmd{In the fingerprint characteristics, \textit{server-version} indicates the preview server version, and \textit{platform} indicates the device platform (Android or iOS). Device name indicates the specific device from which the request was made, e.g., iPhone, iPad.}}
\end{table*}